\documentclass[aps,pra,twocolumn,superscriptaddress,showpacs,twoside]{revtex4}

\usepackage{graphicx}
\usepackage{amsmath}
\usepackage{color}

\newcommand{\D}{\mathrm{d}}
\newcommand{\ds}{\displaystyle}
\newcommand{\rhoML}{\widehat{\rho}_{\textsc{ml}}}
\newcommand{\rhoBM}{\widehat{\rho}_{\textsc{bm}}}
\newcommand{\RML}{\widehat{\cR}_{\textsc{ml}}}
\newcommand{\RSC}{\widehat{\cR}_{\mathrm{sc}}}
\newcommand{\prbDR}[1][\cR]{\mathrm{prob}(D \wedge #1)}

\newcommand{\CD}{\cC_D}
\newcommand{\setC}{\mathbf{C}}
\newcommand{\Exp}[1]{\mathrm{e}^{\mbox{\footnotesize$#1$}}}
\newcommand{\I}{\mathrm{i}}
\newcommand{\tr}[1]{\mathrm{tr}{\left\{#1\right\}}}
\newcommand{\cC}{\mathcal{C}}
\newcommand{\cR}{\mathcal{R}}

\makeatletter
\renewcommand{\Dated@name}{Posted on the arXiv on }
\makeatother

\renewcommand{\surname}[1]{\textsc{#1}}

\begin{document}

\title{Optimal error regions for quantum state estimation}
\date{30 March 2013}

\author{\surname{Shang} Jiangwei}
\affiliation{Centre for Quantum Technologies, %
National University of Singapore, Singapore 117543}

\author{Hui Khoon \surname{Ng}}
\affiliation{Centre for Quantum Technologies, %
National University of Singapore, Singapore 117543}
\affiliation{DSO National Laboratories, Singapore 118230}

\author{Arun \surname{Sehrawat}}
\affiliation{Centre for Quantum Technologies, %
National University of Singapore, Singapore 117543}

\author{\surname{Li} Xikun}
\affiliation{Centre for Quantum Technologies, %
National University of Singapore, Singapore 117543}

\author{Berthold-Georg \surname{Englert}}
\affiliation{Centre for Quantum Technologies, %
National University of Singapore, Singapore 117543}
\affiliation{Department of Physics, %
National University of Singapore, Singapore 117542}

\begin{abstract}
Rather than point estimators, states of a quantum system that
represent one's best guess for the given data, we consider optimal
regions of estimators.
As the natural counterpart of the popular maximum-likelihood point estimator,
we introduce the maximum-likelihood region---the region of largest likelihood
among all regions of the same size.
Here, the size of a region is its prior probability.
Another concept is the smallest credible region---the smallest region 
with pre-chosen posterior probability.
For both optimization problems, the optimal region has constant likelihood on
its boundary. 
We discuss criteria for assigning prior probabilities to regions, and
illustrate the concepts and methods with several examples.
\end{abstract}


\maketitle

\section{Introduction}\label{sec:Intro}
Quantum state estimation (see, for example, Ref.~\cite{LNP649}) is central to
many, if not all, tasks that process quantum information.
The characterization of a source of quantum carriers, the verification of the
properties of a quantum channel, the monitoring of a transmission line used
for quantum key distribution---all three require reliable quantum state
estimation, to name just the most familiar examples.

In the typical situation that we are considering, several independently and
identically prepared quantum-information carriers are measured one-by-one by
an apparatus that realizes a probability-operator measurement (POM), suitably
designed to extract the wanted information.
The POM has a number of outcomes, with detectors that register individual
information carriers (photons in the majority of current experiments), and the
data consist of the observed sequence of detection events 
(``clicks'')~\cite{note:typical}.

The quantum state to be estimated is described by a statistical operator, 
the \emph{state}, and
the data can be used to determine an \emph{estimator} for the state---another 
state that, so one hopes, approximates the actual state well.
There are various strategies for finding such an estimator.
Thanks to the efficient methods that Hradil, \v{R}eh\'a\v{c}ek, and their
collaborators developed for calculating maximum-likelihood estimators
(MLEs, reviewed in Ref.~\cite{MLEreview}; see also Ref.~\cite{TeoYS:thesis}), 
MLEs have become the estimators of choice.
For the given data, the MLE is the state for which the data are more likely
than for any other state.

Since the data have statistical noise, one needs to supplement a point
estimator with error bars of some sort---\emph{error regions}, more generally,
for higher-dimensional problems.
Ad-hoc recipes have been proposed for attaching a vicinity of states to a
given point estimator, often relying on approximations valid only in the limit
of a large amount of data (see Refs.~\cite{Rehacek+2:08} and
\cite{Audenaert+1:09} for examples in quantum state estimation), or involves
resampling of the data (see, for instance, Ref.~\cite{Efron+1:93}). 
By contrast, we wish to use systematic procedures for determining error or
estimator regions from only the data that we \emph{did} observe.

We are, however, not considering estimator regions of any kind,
but specifically \emph{maximum-likelihood regions} (MLRs). 
For the given data, the MLR is that region of pre-chosen size, for which the
data are more likely than any other region of the same size.
The regions referred to here are regions in the space of quantum states
(more precisely: in the reconstruction space; see Sec.~\ref{sec:ReconSp}).
As we shall see, there is an intimate connection between the MLE and the MLRs
for the same data: 
All MLRs contain the MLE, and in the limit of very small size, the MLR is a
small vicinity of the MLE.

The ``size of a region'' is clearly an important notion here.
We agree with Evans, Guttman, and Swartz \cite{Evans+2:06} that,
in the present context of state estimation, it is natural to measure the size
of a region by its prior probability that the actual state lies in the
region, that is: the probability that we assign to the region before any data
are at hand. 
As they should, regions with the same size have the same prior probability;
and the whole state space has unit size $\equiv$ unit prior probability
because the actual state is surely somewhere in the state space.

In addition to MLRs, we also consider \emph{smallest credible regions} (SCRs).
The credibility of a region is its posterior probability, that is: the
probability that the actual state lies in the region, conditioned on the
data (see, for example, Ref.~\cite{Berger:85}).
The SCR, then, is the smallest region with the pre-chosen value of the
credibility. 

It turns out that the problems of finding the MLR and the SCR are duals of
each other. 
Each SCR is also a MLR, and each MLR is a SCR.
In both cases, the optimal regions contain all states for which the 
likelihood of the data exceeds a threshold value. 
In particular, in the limit of small credibility, the SCR is a
small vicinity of the MLE.

The \emph{confidence regions} that were recently studied in the quantum
context by Christandl and Renner \cite{Christandl+1:12}, and by Blume-Kohout
\cite{Blume-Kohout:12}, are markedly different from the SCRs and the MLRs.
Confidence regions give an answer to the following question:
Consider all conceivable data, all sequences of detector clicks that could
possibly be obtained, and assign a region to each sequence;
how do we choose the regions such that a pre-chosen fraction
of the regions (the confidence level) will surely contain the unknown actual
state?
We contrast this with the corresponding question for the SCR:
Consider all permissible states, each a candidate for the unknown actual
state; what is the smallest region, for the observed data, that contains the
actual state with a pre-chosen probability?

The difference between the two questions is simple, yet profound.
When asking for confidence regions, the data are regarded as the random
variable; whereas the observed data are given for the SCR, and the unknown state
is the random quantity.
A further difference to note is that the sizes of the confidence regions play
a minor role in their construction, whereas its size is a crucial property of
a SCR. 
  
Here is a brief outline of the paper.
We set the stage in Sec.~\ref{sec:defs} where we introduce the reconstruction
space, discuss the size of a region, and define the various joint and
conditioned probabilities.
Equipped with these tools, we then formulate in Sec.~\ref{sec:OR} the
optimization problems that identify the MLRs and SCRs and find their
solutions; this is followed by remarks on confidence regions.  
Criteria for choosing unprejudiced priors are the subject of
Sec.~\ref{sec:prior}, and simulated qubit measurements
illustrate the matter in Sec.~\ref{sec:examples}.
We close with an outlook on the problems that need to be solved before MLRs
and SCRs can be computed efficiently for data acquired in actual experiments.

\section{Setting the stage}\label{sec:defs}
\subsection{Reconstruction space}\label{sec:ReconSp}
The $K$ outcomes $\Pi_1$, $\Pi_2$, \ldots, $\Pi_K$ of the POM, with which the
data are acquired, are positive Hilbert-space operators that decompose the
identity, 
\begin{equation}\label{eq:2.1}
  \sum_{k=1}^K\Pi_k=1\quad\mbox{with $\Pi_k\geq0$ for $k=1,2,\ldots,K$.}
\end{equation}
If the state $\rho$ describes the system, 
then the probability $p_k$ that the $k$th detector will click for the next
copy to be measured is
\begin{equation}\label{eq:2.2a}
  p_k=\tr{\Pi_k\rho}=\langle\Pi_k\rangle\,,
\end{equation}
which is the Born rule, of course.
Here, $\rho$ can be any positive operator with unit trace,
\begin{equation}\label{eq:2.3}
  \rho\geq0\,,\quad\tr{\rho}=1\,.
\end{equation}
The positivity of $\rho$ and its normalization ensure the
positivity of the $p_k$s and their normalization
\begin{equation}\label{eq:2.2b}
   p_k\geq0\,,\quad\ds\sum_{k=1}^Kp_k=1\,.  
\end{equation}
Probabilities ${p=(p_1,p_2,\ldots,p_K)}$ for which there is a state $\rho$ such
that Eq.~(\ref{eq:2.2a}) holds, are \emph{permissible} probabilities.
They make up the \emph{probability space}.

The probability space for a $K$-outcome POM is usually smaller than that of a
$K$-sided die because not all positive $p_k$s with unit sum are permitted by
the Born rule.
The quantum nature of the state estimation problem enters \emph{only} in these
additional restrictions on $p$:
Quantum state estimation is standard statistical state estimation with quantum
constraints.
The rich concepts and methods of statistical inference apply immediately to
the quantum situation, modified where necessary to account for the restricted
probability space.

Whereas the $p_k$s are uniquely determined by $\rho$ in accordance with
Eq.~(\ref{eq:2.2a}), the converse is only true if the POM is informationally
complete. 
In any case, there is always a
\emph{reconstruction space} $\cR_0$, a set of $\rho$s that
contains exactly one $\rho$ for each set of permissible
probabilities, consistent with the Born rule.
If there is more than one reconstruction space, it does not matter which one
we choose.
While the probability space is always convex, a convex reconstruction
space may not be available.

The reconstruction space is at most ${(K-1)}$-dimen\-sional, and has a
smaller dimension if fewer probabilities are independent. 
We note that $K$ is always finite, and so is the dimension of the
reconstruction space. 
There are no real-life POMs with an infinite number of outcomes. 

As an example, consider a harmonic oscillator with its infinite-dimensional
state space. 
If the POM has two outcomes with $p_1$ equal to the probability of finding the
oscillator in its ground state, and $p_2=1-p_1$, one reconstruction space
is the set of convex combinations of the projector to the ground state and
another state with no ground-state component.
In this situation, there is a large variety of reconstruction spaces to choose
from, because \emph{any} other state serves the purpose, 
and all one can infer from the data is an estimate of the
ground-state probability.  

Now, state estimation is the task of finding a state, or a region of states,
in the reconstruction space by a systematic and reliable procedure that
exploits the observed data.
In view of the one-to-one correspondence between the states in the
reconstruction space and the permissible probabilities, one can identify
the reconstruction space with the probability space.
Indeed, since the probability space is unique, 
while there can be many different reconstruction spaces, it is often more
convenient to work in the probability space.
The primary objective is then to find an estimator, or a region of estimators,
for the probabilities $p$.
The conversion of the set of probabilities $p$ into a state $\rho$ is performed
later, if at all, and only at this stage do we need to decide which
reconstruction space is used for reference. 
If the POM is not informationally complete, it will be necessary to
invoke additional criteria or principles for a unique mapping ${p\to\rho}$. 
For example, one could follow Jaynes's guidance \cite{Jaynes:57a,Jaynes:57b}
and maximize the entropy~\cite{Teo+4:11} (see also Ref.~\cite{MaxEntreview}).

\subsection{Size and prior content of a region}\label{sec:SizePrior}
Prior to acquiring any data, we assign equal probabilities to equivalent
alternatives. 
If we split the reconstruction space in two, it is equally likely that the
actual state is in either half and, therefore, each half should carry a prior
probability of $\frac{1}{2}$, provided that the splitting-in-two is fair, 
that is: the two pieces are of equal size.
A preconceived notion of size is taken for granted here.
Further fair splitting, into more disjoint regions of equal size, then suggests
rather strongly that the prior probability of a region should be proportional
to its size.
We take this suggestion seriously: Scale all region sizes such that the whole
reconstruction space has unit size, and then the size of a region \emph{is}
its prior probability---its ``prior content'' if we borrow terminology from
Bayesian statistics.

The identification ``size $\equiv$ prior probability'' is technically possible
because both quantities simply add if disjoint regions are combined into a
single region. 
There is no room for mathematical inconsistencies here, unless we begin with
a region-to-size mapping for which the reconstruction space cannot be
normalized to unit size, so that we would obtain improper prior probabilities.
We are not interested in pathological cases of this or other kinds 
and just exclude them.
Should an improper prior be useful in a particular context, it should come
about as the limit of a well-defined sequence of proper priors.

The above line of reasoning can be reversed.
Should we have established each region's prior probability with other means
(perhaps invoking symmetry arguments or taking into account that the source
under investigation is designed to emit the information carriers in a
certain target state; see Sec.~\ref{sec:prior}), then we accept this as the
natural measure of the region's size~\cite{Evans+2:06}.
After all, the reconstruction space is an abstract construct that is not
endowed with a self-suggesting unique metric, and a region's prior probability
is the quantity that matters most in the present context of statistical
inference. 

We denote by $(\D\rho)$ the size of the infinitesimal vicinity of state
$\rho$.
The size $S_{\cR}$ of a region $\cR\subseteq\cR_0$ is then obtained by
integrating over the region,
\begin{equation}\label{eq:2.4}
  S_{\cR}=\int\limits_{\cR}(\D\rho)
\quad\mbox{with $\ds\int\limits_{\cR_0}(\D\rho)=1$}\,,
\end{equation}
where the latter integration covers all of the reconstruction space.
By construction, the value of $S_{\cR}$ does not depend on the
parameterization that we use for the numerical representation of $(\D\rho)$.
The primary parameterization is in terms of the probabilities,
\begin{equation}\label{eq:2.5}
  (\D\rho)=(\D p)\,w(p)
\quad\mbox{with $\ds(\D p)=\D p_1\,\D p_2\,\cdots\,\D p_K$}\,,
\end{equation}
where the prior density $w(p)$ is nonzero for all permissible probabilities
and vanishes for all non-permissible $p$s.
In particular, $w(p)$ always contains 
\begin{equation}\label{eq:2.6}
  w_0(p)=\eta(p_1)\eta(p_2)\cdots\eta(p_K)\delta(p_1+p_2+\cdots+p_K-1)
\end{equation}
as a factor and so enforces the constraints that the probabilities are
positive and have unit sum~\cite{note:eta-delta}.
If there are no other constraints, we have the probability space of a
$K$-sided die.
For genuine quantum measurements, however, there are additional constraints, 
some accounted for by more delta-function factors, others by step functions.
The delta-function constraints reduce the dimension of the
reconstruction space from ${K-1}$ to the number of independent probabilities.

For the harmonic-oscillator example of Sec.~\ref{sec:ReconSp},
which has the same probability space as a tossed coin, 
the factor $w_0(p)$ selects the line segment with
${0\leq p_1=1-p_2\leq1}$ in the $p_1p_2$ plane. 
If we choose the ``primitive prior'' ${(\D\rho)=(\D p)\,w_0(p)}$, 
the subsegment with ${a\leq p_1\leq b}$ has size ${b-a}$.
For the Jeffreys prior~\cite{Jeffreys:46}, a popular choice of an
  unprejudiced prior~\cite{Kass+1:96},
\begin{eqnarray}\label{eq:2.6a}
  (\D\rho)=(\D p)\,w_0(p)\frac{1}{\pi\sqrt{p_1p_2}}\,,
\end{eqnarray}
the same subsegment has size 
$\frac{2}{\pi}[\sin^{-1}(\sqrt{b})-\sin^{-1}(\sqrt{a})]$.

In this example, and also in those we use for illustration in
Sec.~\ref{sec:examples} below, it is easy to state quite explicitly the
restrictions on the set of permissible probabilities that follow from the Born
rule. 
In other situations, it could be difficult or impossible.
This is why state estimation is often done by searching for a statistical
operator in a suitable state space.
For practical reasons, it may be necessary to truncate the full state
space---which can be, and often is, infinite-dimensional---to a test space of
manageable size.
With such a truncation one accepts that not all permissible probabilities are
investigated.
Therefore, a criterion for judging if the test space is large enough is to
verify that the estimated probabilities do not change significantly when the
space is enlarged.
Examples for the artifacts that result from test spaces that are too small can
be found in Ref.~\cite{Teo+4:12}.

\subsection{Point likelihood, region likelihood, credibility}
\label{sec:Lik}
The data $D$ acquired by the POM consist of a sequence of detector clicks, 
with a total of
$n_k$ clicks of the $k$th detector, and a total number of
$N=n_1+n_2+\cdots+n_K$ clicks after measuring $N$ quantum-information
carriers~\cite{note:imperfections}. 
The probability of obtaining the data, \emph{if} $\rho$ is the state, is the
familiar \emph{point likelihood} 
\begin{equation}\label{eq:2.7}
  L(D|\rho)=p_1^{n_1}p_2^{n_2}\cdots p_K^{n_K}\,.
\end{equation}
It attains its maximal value when $\rho$ is the MLE
$\rhoML$,
\begin{equation}\label{eq:2.8}
  \max\limits_{\rho}L(D|\rho)=L(D|\rhoML)\,,
\end{equation}
where $\rhoML$ is in the reconstruction space, but the maximum could be taken
over all states.

The joint probability of finding the state $\rho$ in the region $\cR$ and
obtaining the data $D$ is then
\begin{equation}\label{eq:2.9}
  \prbDR=\int\limits_{\cR}(\D\rho)\,L(D|\rho)\,.
\end{equation}
If ${\cR=\cR_0}$, we have the prior likelihood
$L(D)$,
\begin{equation}\label{eq:2.10}
  \prbDR[\cR_0]=L(D)=\int\limits_{\cR_0}(\D\rho)\,L(D|\rho)\,.
\end{equation}
Since one of the click sequences is surely observed, the likelihoods of
Eqs.~(\ref{eq:2.7}) and (\ref{eq:2.10}) have unit sum,
\begin{eqnarray}\label{eq:2.11}
  \sum_DL(D|\rho)&=&\sum_{n_1,\dots,n_K}\!\!\frac{N!\,\delta_{N,n_1+n_2+\cdots+n_K}}
{n_1!\,n_2!\,\cdots\,n_K!}
p_1^{n_1}p_2^{n_2}\cdots p_K^{n_K}\nonumber\\
&=&(p_1+p_2+\cdots+p_K)^N=1\,,\nonumber\\
\sum_DL(D)&=&\int\limits_{\cR_0}(\D\rho)=1\,.
\end{eqnarray}

We factor the joint probability $\prbDR$ in two different ways,
\begin{equation}\label{eq:2.12}
  \prbDR=L(D|\cR)S_{\cR}=C_{\cR}(D)L(D)\,,
\end{equation}
and so identify the \emph{region likelihood} $L(D|\cR)$ and the 
\emph{credibility} $C_{\cR}(D)$.
Both quantities are conditional probabilities:
The region likelihood is the probability of obtaining the data $D$ if the
state is in the region $\cR$;
the credibility is the probability that the actual state is in the region
$\cR$ if the data $D$ were obtained---the posterior probability of $\cR$.

\section{Optimal error regions}\label{sec:OR}
\subsection{Maximum-likelihood regions}\label{sec:MLR}
Instead of looking for the MLE, the single point in the reconstruction space
that has the largest likelihood for the given data $D$, we desire a region with
the largest likelihood---the MLR.
For this purpose, we maximize the region likelihood $L(D|\cR)$ under the
constraint that only regions with a pre-chosen size $s$ participate in the
competition, with ${0<s<1}$; an unconstrained maximization of $L(D|\cR)$ is
not meaningful because it gives the limiting region that consists of nothing
but the point $\rhoML$. 
The resulting MLR $\RML$ is a function of the data $D$ and the size $s$, but
we wish to not overload the notation and will keep these dependences implicit,
just like the notation does not explictly indicate the $D$ dependence of the
MLE $\rhoML$.

The MLR analog of the MLE definition in Eq.~(\ref{eq:2.8}) is then
\begin{equation}\label{eq:3.1}
  \max_{\cR\subseteq\cR_0}L(D|\cR)=L(D|\RML)\quad\mbox{with $S_\cR=s$}\,.
\end{equation}
Since all competing regions have the same size, we can equivalently maximize
the joint probability,
\begin{equation}\label{eq:3.2}
  \max_{\cR\subseteq\cR_0}\prbDR=\prbDR[\RML]\quad\mbox{with $S_\cR=s$}\,.
\end{equation}
The answer to this maximization problem is given in Corollary~4 of
Ref.~\cite{Evans+2:06} and justified by a detailed proof of considerable
mathematical sophistication.
We proceed to offer an alternative argument that is perhaps more accessible
to the working physicist.

\begin{figure}
\centering
\includegraphics{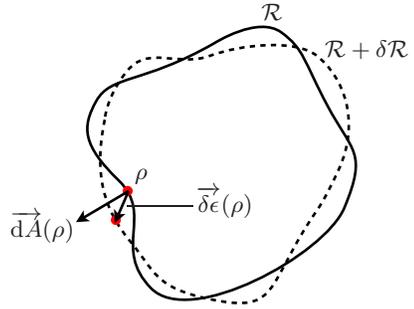}
\caption{\label{fig:RegionVary}%
Infinitesimal variation of region $\cR$. The boundary of region $\cR$ (solid
line) is deformed to become the boundary of region $\cR+\delta\cR$ (dashed
line).} 
\end{figure}

Owing to the maximum property of the MLR and its fixed size, both $\prbDR$ and
$S_\cR$ must be stationary under infinitesimal variations $\delta\cR$
of the region $\cR$. 
 Such an infinitesimal variation is achieved by deforming the boundary
$\partial\cR$ of the region, as illustrated in Fig.~\ref{fig:RegionVary}.
The resulting change in the size $S_\cR$ vanishes for all permissible
deformations,
\begin{equation}\label{eq:3.3}
\delta S_{\cR} = \int\limits_{\partial\cR} 
\overrightarrow{\D A}(\rho)\cdot\overrightarrow{\delta\epsilon}(\rho)=0\,.  
\end{equation}
Here, $\overrightarrow{\D A}(\rho)$ is the vectorial surface element of the
boundary $\partial\cR$ at point $\rho$ in the reconstruction space, and
$\overrightarrow{\delta\epsilon}(\rho)$ is the infinitesimal displacement of
the point $\rho$ that deforms $\cR$ into $\cR+\delta\cR$. 

\begin{figure}
\centering
\includegraphics{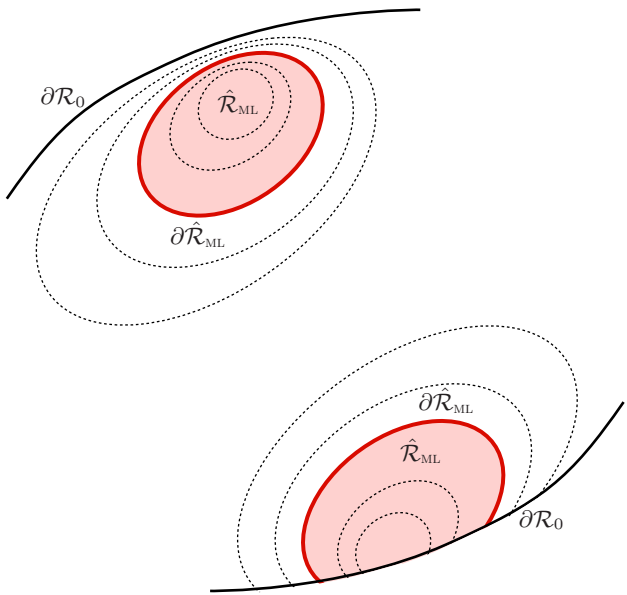}
\caption{\label{fig:BoundaryR} MLRs of two kinds. In the top-left sketch,
  $\RML$ is completely contained inside the reconstruction space; in the
  bottom-right   sketch, the boundary $\partial\RML$ of $\RML$ contains 
  a part of the surface $\partial\cR_0$ of the reconstruction space.  
  Dotted lines indicate iso-likelihood surfaces, 
  that is: surfaces on which the point likelihood is constant.}
\end{figure}

The corresponding change in $\prbDR$ is
\begin{equation}\label{eq:3.4}
  \delta\prbDR=\int\limits_{\partial\cR} 
\overrightarrow{\D A}(\rho)\cdot\overrightarrow{\delta\epsilon}(\rho)
\,L(D|\rho)=0\,,
\end{equation}
which attains the indicated value of $0$ at the extremum $\cR=\RML$.
If we have the situation sketched in the top-left plot of
Fig.~\ref{fig:BoundaryR}, where $\RML$ is completely in the interior of the
reconstruction space,  
both Eqs.~\eqref{eq:3.3} and \eqref{eq:3.4} must hold simultaneously for
arbitrary infinitesimal deformation $\delta\cR$. 
This is possible only if the point likelihood $L(D|\rho)$ is constant on the
boundary $\partial\RML$ of $\RML$, that is: $\partial\RML$ is an
\emph{iso-likelihood surface} (ILS). 
Furthermore, $\RML$ must correspond to the interior of this
ILS (as opposed to its complement in the reconstruction
space), since the concavity of the logarithm of the point likelihood implies
that the interior necessarily has larger likelihood values than its
complement~\cite{note:concave}.

If the boundary $\partial\RML$ of $\RML$ contains a part of the surface
$\partial\cR_0$ of the reconstruction space, which is the situation on
the bottom-right in Fig.~\ref{fig:BoundaryR}, all interior points on 
$\partial\RML$ must still lie on an ILS, or else we can
always deform $\partial\RML$ to attain a larger value of the region likelihood
with a permissible choice of $\overrightarrow{\delta\epsilon}(\rho)$.  
On the $\partial\cR_0$ part of $\partial\RML$, the point likelihood
$L(D|\rho)$ has larger values than the constant value on the interior part of
the boundary, because ILSs that are inside $\RML$ (dashed in
Fig.~\ref{fig:BoundaryR}) and have endpoints in $\partial\cR_0$ assign their
larger likelihood values to these points.
Therefore, deforming the $\partial\cR_0$ part of $\partial\RML$ inwards, with
the change in size compensated for by an outwards deformation of the interior
part of $\partial\RML$, decreases the value of the region likelihood.
And since outwards deformations of $\partial\cR_0$ are not possible, a region
with an ILS as interior part of the boundary, supplemented by a
part of $\partial\cR_0$, is a possible MLR, indeed.

In summary, the MLRs of various sizes $s$ consist of all states $\rho$ for
which the point likelihood $L(D|\rho)$ exceeds a certain threshold value,
with higher thresholds for smaller sizes. 
Quite remarkably and somewhat surprisingly, the set of MLRs does not depend on
the chosen prior. 
The shape of a MLR is fully determined by the point likelihood and the
threshold value; the prior enters only when the size, region likelihood, and
credibility of the MLR are calculated.

It is expedient to specify the threshold value as a fraction of the maximum
value $L(D|\rhoML)$ of the point likelihood. 
Denoting this fraction by $\lambda$, the characteristic function of the
corresponding \emph{bounded-likelihood region} (BLR) $\cR_\lambda$ is the step
function 
\begin{equation}\label{eq:3.5}
\eta^{\ }_{\cR_\lambda}\!(\rho)
= \eta\bigl(L(D|\rho)-\lambda L(D|\rhoML)\bigr)\,,  
\end{equation}
where
\begin{equation}\label{eq:3.5'}
\eta_\cR^{\ }(\rho)
=\left\{\begin{array}{ll}
    1&\mbox{if $\rho$ is in $\cR$}\\
    0&\mbox{else}
  \end{array}\right.
\end{equation}
is the characteristic function of region $\cR$.
BLRs have appeared previously in standard statistical analysis; see
Ref.~\cite{Wasserman:89} and references therein.

The BLR $\cR_\lambda$ has the size
\begin{equation}\label{eq:3.6}
  s_\lambda=\int\limits_{\cR_0}(\D\rho)\,\eta^{\ }_{\cR_\lambda}\!(\rho)\,,
\end{equation}
and we have ${\cR_\lambda=\cR_0}$ and ${s_\lambda=s_0=1}$ for
${\lambda\leq\lambda_0}$ with ${\lambda_0\geq0}$
given by
\begin{equation}\label{eq:3.6'}
   \min\limits_\rho L(D|\rho)=\lambda_0\, L(D|\rhoML)\,.
\end{equation}
As $\lambda$ increases from $\lambda_0$ to $1$, $s_\lambda$
decreases monotonically from $1$ to $0$.
The size $s$ specified in Eq.~(\ref{eq:3.1}) is obtained for an intermediate
$\lambda$ value, and the corresponding BLR is the looked-for MLR.

The MLE is contained in all MLRs.
In the ${s\to0}$ limit, the MLR becomes an infinitesimal vicinity of the MLE
and the region likelihood of the limit region is equal to the point likelihood
of the MLE, ${L(D|\RML)\to L(D|\rhoML)}$.

\subsection{Smallest credible regions}\label{sec:SCR}
The MLR is the region for which the observed data are particularly likely.
With a reversal of emphasis, we now look for a region that contains the actual
state with high probability.
Ultimately, this is the SCR $\RSC$: the smallest region for which the
credibility has the pre-chosen value~$c$.

For the given $D$, the optimization problem
\begin{equation}\label{eq:3.7}
  \min_{\cR\subseteq\cR_0} S_{\cR}=S_{\RSC}\quad
\mbox{with $C_{\cR}(D)=c$}
\end{equation}
is dual to that of Eqs.~(\ref{eq:3.1}) and (\ref{eq:3.2}).
Here we minimize the size for given joint probability, there we maximize the
joint probability for given size.
It follows that the BLRs of Eq.~(\ref{eq:3.5}) are not only the MLRs, they are
also the SCRs: Each MLR is a SCR, each SCR is a MLR.

The BLR $\cR_\lambda$ has the credibility
\begin{equation}
  \label{eq:3.8}
  c_\lambda=\frac{1}{L(D)}\int\limits_{\cR_0}(\D\rho)
           \,\eta^{\ }_{\cR_\lambda}\!(\rho)L(D|\rho)\,,
\end{equation}
which, just like $s_\lambda$, decreases monotonically from $1$ to~$0$ as
$\lambda$ increases from $\lambda_0$ to $1$.
The credibility $c$ specified in Eq.~(\ref{eq:3.7}) is obtained for an
intermediate value, and the corresponding BLR is the looked-for SCR.

\subsection{Size and credibility of a BLR}\label{sec:scBLR}
The responses of the size $s_\lambda$ and the credibility $c_\lambda$ of a BLR
to an infinitesimal change of $\lambda$ are linked by
\begin{equation}\label{eq:3.12}
  L(D)\frac{\partial}{\partial\lambda}c_\lambda=
  L(D|\rhoML)\lambda\frac{\partial}{\partial\lambda}s_\lambda\,.
\end{equation}
Therefore, once $s_\lambda$ is known as a function of $\lambda$, we obtain
$c_\lambda$ by an integration,
\begin{equation}
  \label{eq:3.12a}
  c_{\lambda}=\frac{\ds\lambda s_\lambda+
                \mathop{\mbox{\small$\ds\int_\lambda^1$}}\D\lambda'\,s_{\lambda'}}
    {\ds\mathop{\mbox{\small$\ds\int_0^1$}}\D\lambda'\,s_{\lambda'}}\,.
\end{equation}
This is, of course, consistent with the limiting values for 
${\lambda\leq\lambda_0}$ and
${\lambda=1}$, and also establishes that, for all intermediate values,
the credibility of a BLR is
larger than its size,
\begin{equation}\label{eq:3.9}
  c_\lambda>s_\lambda\quad\mbox{for ${\lambda_0<\lambda<1}$}\,.
\end{equation}
Further, Eqs.~(\ref{eq:3.12}) and (\ref{eq:3.12a}) tell us that in the
${\lambda\to1}$ limit, when both $s_\lambda$ and $c_\lambda$ vanish, their
ratio is finite and exceeds unity, 
\begin{equation}\label{eq:3.13}
  \frac{c_\lambda}{s_\lambda}\to
\frac{1}{\ds\mathop{\mbox{\small$\ds\int_0^1$}}\D\lambda'\,s_{\lambda'}}=
\frac{L(D|\rhoML)}{L(D)}>1
\quad\mbox{for $\lambda\to1$}\,.
\end{equation}
We note that this provides the value of $L(D)$, since the maximal value
$L(D|\rhoML)$ of the point likelihood is computed earlier as it is needed 
for identifying the BLRs. 

Inasmuch as the value of $s_\lambda$ quantifies our prior belief that the
actual state is in $\cR_\lambda$, we are surprised when the data tell us that
the probability for finding the state in that region is larger. 
Accordingly, the SCR is the region for which we are most 
surprised for the given prior belief~\cite{note:evidence}.
This matter and other aspects of Bayesian inference based on the concept of
relative surprise are discussed in Ref.~\cite{Evans+2:06}.   

The relation (\ref{eq:3.12a}) is also of considerable practical importance
because we only need to evaluate the multi-dimensional integrals of
Eq.~(\ref{eq:3.6}), but not those of Eqs.~(\ref{eq:3.8}) and (\ref{eq:2.10}).
Since the latter integrals require well-tailored Monte-Carlo methods to handle
the typically sharply peaked likelihood function, the
numerical effort is very substantially reduced if we only need to evaluate the
integral of Eq.~(\ref{eq:3.6}).

Indeed, the estimator regions for the observed data are conveniently and
concisely communicated by reporting $s_\lambda$ and $c_\lambda$ as functions
of $\lambda$.
The end users interested in the MLR with the size of his liking or the SCR of
her wanted credibility can thus determine the corresponding values of
$\lambda$.
It is then an easy matter to check if any particular $\rho$ is inside the
specified region or not.

Once more, we use the simple harmonic-oscillator example of
Sec.~\ref{sec:ReconSp} for illustration.
Suppose, $N=2$ copies have been measured, and we obtained one click each for
the two outcomes, so that the point likelihood is equal to $p_1p_2$.
In this situation, we have ${\lambda_0=0}$ and 
$\eta^{\ }_{\cR_\lambda}\!(\rho)=\eta(4p_1p_2-\lambda)$, so that
$\bigl|p_1-p_2\bigr|\leq\sqrt{1-\lambda}$ for the BLR $\cR_\lambda$.
This gives
\begin{eqnarray}\label{eq:3.14}
  s_\lambda&=&\sqrt{1-\lambda}\,,
\nonumber\\
  c_\lambda&=&\frac{1}{2}(2+\lambda)\sqrt{1-\lambda}
\end{eqnarray}
for the primitive prior, and 
\begin{eqnarray}\label{eq:3.15}
  s_\lambda&=&1-\frac{2}{\pi}\sin^{-1}(\sqrt{\lambda})\,,
\nonumber\\
  c_\lambda&=&1-\frac{2}{\pi}\sin^{-1}(\sqrt{\lambda})
            +\frac{2}{\pi}\sqrt{\lambda(1-\lambda)}
\end{eqnarray}
for the Jeffreys prior.

\subsection{Confidence regions}\label{sec:ConfR}
The confidence regions that were recently studied by Christandl and Renner
\cite{Christandl+1:12}, and independently by Blume-Kohout
\cite{Blume-Kohout:12}, are markedly different from the MLRs and the SCRs.
The MLR and the SCR represent inferences drawn about the unknown state $\rho$
from the data $D$ that have actually been observed.
By contrast, confidence regions are a set of regions, one region for each
data, whether observed or not, from the
measurement of $N$ copies. 
The confidence regions would contain \emph{any} state in, at least, a certain
fraction of many $N$-copy measurements, if the many measurements were 
performed.  
This fraction is the confidence level.

When denoting by $\CD$ the confidence region for data $D$, the confidence level
$\gamma$ of the set $\setC$ of $\CD$s for all conceivable data (for fixed $N$)
is 
\begin{equation}\label{eq:3.16}
  \gamma(\setC)=\min_{\rho}\sum_D L(D|\rho) \,\eta_{\CD}^{\ }\!(\rho)\,,
\end{equation}
where the minimum is reached in the ``worst case.''
For example, in the security analysis of a protocol for quantum key
distribution, one wishes a large value of $\gamma$ to protect against an
adversary who controls the source and prepares the quantum-information
carriers in the state that is best for her.

Any set $\setC$, for which $\gamma$ has the desired value, serves the purpose.
A smaller set $\setC'$, in the sense that $\CD'$ is contained in $\CD$ for all
$D$, is preferable, but usually there is no smallest set of confidence regions.
Here, ``smaller'' is solely in this inclusion sense, with no reference to a
quantification of the size of a region and, therefore, there is no necessity
of specifying the prior probability of any region.
Since the transition from set $\setC$ to the smaller set $\setC'$ requires the
shrinking of some of the $\CD$s without enlarging even a single one,
it is easily possible to have two sets of confidence regions with the same
confidence level and neither set smaller than the other.

\begin{figure}
\centering
\includegraphics{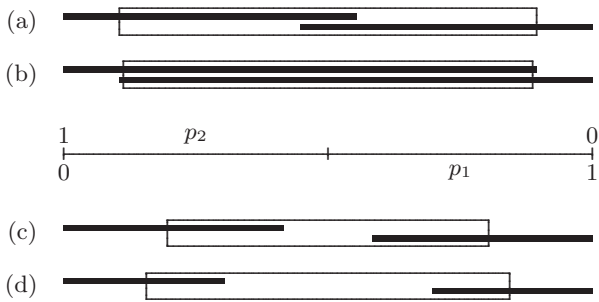}
\caption{\label{fig:confidence}
Confidence regions and smallest credible regions.
The bars indicate intervals of ${p_1=1-p_2}$ for the harmonic-oscillator
example of Sec.~\ref{sec:ReconSp}, which has the reconstruction space of a
tossed coin.
Two copies are measured.
The left solid bars indicate the regions for $(n_1,n_2)=(0,2)$ counts; the
right solid bars are for $(n_1,n_2)=(2,0)$; and the central open bars are for
$(n_1,n_2)=(1,1)$. 
Cases (a) and (b) show two sets of confidence regions for confidence level
${\gamma=0.8}$.
Regions (c) and (d) are the SCRs for the primitive prior and the
Jeffreys prior, respectively, both for credibility ${c=0.8}$.
}
\end{figure}

For illustration, we consider the harmonic-oscillator example of
Sec.~\ref{sec:ReconSp} yet another time.
Figure \ref{fig:confidence} shows two sets of confidence regions
(${\gamma=0.8}$) and the corresponding three SCRs (${c=0.8}$) for
the primitive prior and the Jeffreys prior.
Both sets of confidence regions are optimal in the sense that one cannot
shrink even one of the regions without decreasing the confidence level, but
neither set is smaller than the other.
In the absence of additional criteria that specify a preference, both work
equally well as sets of confidence regions. 

We observe in this example that confidence regions tend to overlap a lot,
which is indeed unavoidable if a large confidence level is desired.
By contrast, the SCRs for different data usually do not overlap
unless the data are quite similar.
In Fig.~\ref{fig:confidence}, there is no overlap of the SCRs for
${(n_1,n_2)=(0,2)}$ and $(2,0)$.

An important difference of considerable concern in all practical applications
is the following.
Once the data are obtained, there is \emph{the} MLR and \emph{the} SCR for
these data, and it plays no role what other MLRs or SCRs are associated with
different data that have not been observed.
To find the confidence region for the actual data, however, one must first
specify the whole set $\setC$ of confidence regions because the confidence
level of Eq.~(\ref{eq:3.16}) is a property of the whole set.
Christandl and Renner~\cite{Christandl+1:12} have shown that one can choose
high-credibility regions for the $\CD$s~\cite{note:SCRs for CDs}, and
Blume-Kohout~\cite{Blume-Kohout:12} has argued that a set $\setC$ composed of
BLRs can be a pretty good set of confidence regions.

\section{Choosing the prior}\label{sec:prior}
The assignment of prior probabilities to regions in the reconstruction space
should be done in an unprejudiced manner while taking into account all prior
information that might be available.   
We cannot do justice to the rich literature on this subject and
are content with noting that Ref.~\cite{Kass+1:96} reviews various approaches
to constructing unprejudiced priors. 
Let us discuss some criteria that are useful when choosing a prior. 

A general remark is this: 
The chosen prior should give some weight to (almost) all states, and it should
not give extremely high weight to states in some part of the state space
and extremely low weight to other states. 
This is to say that the prior should be \emph{consistent} in the sense that
the credibility of a region---its posterior content---is dominated by the data,
rather than by the prior, if a reasonably large number $N$ of copies is
measured.

\subsection{Uniformity}\label{sec:uniform}
The time-honored strategy of choosing a uniform prior gets us into a circular
argument:
The line of thought presented in Sec.~\ref{sec:SizePrior} implements this
strategy and leads to identifying the prior content of a region with its size.
But that just means that we are now asked to declare how we measure the size
of a region without prejudice, which is the original question about the prior.

In fact, there is no unique meaning of the uniformity of a prior.
In the sense that each prior tells us how to quantify the size of a region,
each prior is uniform with respect to its induced size measure.

This point can be illustrated with the harmonic-oscillator example
of Sec.~\ref{sec:ReconSp}.
For the primitive prior of Sec.~\ref{sec:SizePrior}, the parameterization
\begin{eqnarray}\label{eq:4.1}
  &&p_1=\frac{1}{2}(y+x)\,,\quad p_2=\frac{1}{2}(y-x)\,,\nonumber\\
&&\D p_1\,\D p_2=\D x\,\D y\,\frac{1}{2}
\end{eqnarray}
gives
\begin{eqnarray}\label{eq:4.2}
(\D\rho)&=& \D x\,\D y\,\frac{1}{2}\eta(y+x)\eta(y-x)\delta(y-1)\nonumber\\
&\to&\D x\,\frac{1}{2}\quad\mbox{with $-1\leq x\leq1$}\,,  
\end{eqnarray}
where we integrate over $y$ in the last step and so observe that the primitive
prior is uniform in $x$, that is: the size of the region ${x_1<x<x_2}$ is
proportional to ${x_2-x_1}$.
Likewise, the parameterization
\begin{eqnarray}\label{eq:4.3}
  &&p_1=y(\sin\alpha)^2\,,\quad
    p_2=y(\cos\alpha)^2\,,\nonumber\\
  &&\D p_1\,\D p_2=\D\alpha\,\D y\,y\sin(2\alpha)
\end{eqnarray}
gives
\begin{equation}\label{eq:4.4}
  (\D\rho)\to\D\alpha\,\frac{2}{\pi}\quad
\mbox{with $\ds0\leq\alpha\leq\frac{\pi}{2}$}
\end{equation}
for the Jeffreys prior, which is uniform in $\alpha$.
Other priors can be treated analogously, each of them yielding a uniform prior
in an appropriate single parameter.

The parameterizations in Eqs.~(\ref{eq:4.1}) and (\ref{eq:4.3}) exhibit in
which explicit sense the primitive prior and the Jeffreys prior are uniform.
But the priors are what they are, irrespective of how they are parameterized.
They are explicitly uniform in a particular parameterization and implicitly
uniform in all others.
Uniformity, it follows, cannot serve as a principle that distinguishes one
prior from another.

This ubiquity of uniform priors for a continuous set of infinitesimal
probabilities is in marked contrast to situations in which prior probabilities
are assigned to a finite number of discrete possibilities, such as
the $38$ pockets of a double-zero roulette wheel.
Uniform probabilities of $1/38$ suggest themselves, are meaningful, and
clearly distinguished from other priors, all of which have a bias.

Uniformity in a particularly natural parameterization of the probability space
might also be meaningful.
This, however, invokes a notion of ``natural'' that others may not share.

\subsection{Utility}\label{sec:utility}
In many applications, estimating the state is not a purpose in itself, but only
an intermediate step on the way to determining some particular property of the
physical system.
The objective is to find the value of a parameter that quantifies the
\emph{utility} of the state. 

For example, one could be interested in the fidelity of the actual state with a
target state, or in an entanglement measure of a two-partite state, or in
another quantity that tells us how useful are the quantum-information carriers
for their intended task.
In a situation of this kind, one should, if possible, use a prior that is
uniform in the utility parameter of interest.

As a simple example, consider a single qubit.
The utility parameter is the purity $\xi(\rho)=\tr{\rho^2}$ of the 
state $\rho$.
With the Bloch-ball representation of a qubit state, 
$\rho=\frac{1}{2}(1+\boldsymbol{r}\cdot\boldsymbol{\sigma})$, where
${\boldsymbol{r}=\tr{\boldsymbol{\sigma}\rho}%
=\langle\boldsymbol{\sigma}\rangle}$ 
is the Bloch vector and $\boldsymbol{\sigma}$ is the vector of Pauli
matrices, the purity is 
\begin{equation}
\xi(\rho)=\frac{1}{2}(1+r^2)\quad\mbox{with $r=\boldsymbol{|r|}$}
\,.
\end{equation}

A prior uniform in purity induces a prior on the state space according to
\begin{equation}\label{eq:4.x}
(\D\rho)\propto\D\xi\,\D\Omega\propto r\D r\,\D\Omega,
\end{equation}
where we parameterize the Bloch ball by spherical 
coordinates $(r,\theta,\phi)$.
Here, $\D\Omega$ is the prior for the angular coordinates; 
the prior for the radial coordinate $r$ is fixed by our choice of uniformity 
in $\xi$.
Irrespective of what we choose for $\D\Omega$, the marginal prior for $r$ is
uniform in $\xi$.

If one can quantify the utility of an estimator by a cost function, an optimal
prior can be selected by a minimax strategy:
For each prior in the competition one determines the maximum of the cost
function over the states in the reconstruction space, and then chooses the
prior for which the maximum cost is minimal.
In classical statistics, such minimax strategies are common (see, for instance,
Chapter~5 in Ref.~\cite{Lehmann+1:98});
for an example in the context of quantum state estimation, 
see Ref.~\cite{Ng+2:12}.

\subsection{Symmetry}\label{sec:symmetric}
Symmetry considerations are often helpful in narrowing the search for the
appropriate prior.
For a particularly instructive example, see Sec.~12.4.4 in Jaynes's
posthumous book~\cite{Jaynes:03}.  

Returning to the uniform-in-purity prior of Eq.~(\ref{eq:4.x}), one can
invoke rotational symmetry in favor of the usual solid-angle element, 
$\D\Omega=\sin\theta\D\theta\,\D\phi$, as the choice of angular prior.
The reasoning is as follows: 
The purity of a qubit state does not change under unitary transformations;
unitarily equivalent states have the same purity.
Now, regions that are turned into each other by a unitary transformation have
identical radial content whereas the angular dependences are related by a
rotation.
Invariance under rotations, in turn, requires that the prior is proportional
to the solid angle, hence the identification of $\D\Omega$ with the
differential of the solid angle.  
Note that the resulting prior element $(\D\rho)$ is different from the usual
Euclidean volume element, $r^2\D r\,\sin\theta\D\theta\,\D\phi$, which would be
natural if the Bloch ball were an object in the physical three-dimensional
space. 
But it ain't. 

Symmetry arguments should be used carefully and not blindly.
For a fairly tossed coin, the prior should not be affected if the
probabilities for heads and tails are interchanged, ${w(p_1,p_2)=w(p_2,p_1)}$.
However, for the harmonic-oscillator example of Sec.~\ref{sec:ReconSp}, which
has the same reconstruction space as the coin, there is poor justification for
requiring this symmetry because the two probabilities---of finding the
oscillator in its ground state, or not---are not on equal footing.

\subsection{Invariance}\label{sec:invariant}
When one speaks of an \emph{invariant prior}, one does not mean the invariance
under a change of parameterization\linebreak[0]---all priors are invariant in
this respect---but rather a \emph{form-invariant} construction in terms of a
quantity that, preferably, has an invariant significance.
We consider two particular constructions that make use of the metric induced
by the response of the selected function to infinitesimal changes of its
variables. 

The first construction begins with a quantity $F(p)$ that is a function of all
probabilities ${p=(p_1,\ldots,p_K)}$.
We include the square root of the determinant of the dyadic second derivative
in the prior density as a factor,
\begin{equation}\label{eq:4.5}
  (\D\rho)=(\D p)\,\Bigglb|\det{\left\{{\left(
            \frac{\partial^2F}{\partial p_j\,\partial p_k}
            \right)}_{jk}\right\}}\Biggrb|^{1/2}
            w_{\mathrm{cstr}}(p)\,,
\end{equation}
where $w_{\mathrm{cstr}}(p)$ contains all the delta-function and step-function
factors of constraint as well as the normalization factor that ensures the
unit size of the reconstruction space~\cite{note:independence}.
The prior defined by Eq.~(\ref{eq:4.5}) is invariant in the sense that a
change of parameterization, from $p$ to $\alpha$, say, does not affect its
structure,
\begin{equation}\label{eq:4.6}
  (\D\rho)=(\D\alpha)\,\Bigglb|\det{\left\{{\left(
            \frac{\partial^2F}{\partial\alpha_j\,\partial\alpha_k}
            \right)}_{jk}\right\}}\Biggrb|^{1/2}
            w_{\mathrm{cstr}}\bigl(p(\alpha)\bigr)\,,  
\end{equation}
because the various Jacobian determinants take care of each other.

For the second construction, we use a data-dependent function $G(p,\nu)$ of
the probabilities $p$ and the frequencies
${\nu=(\nu_1,\nu_2,\ldots,\nu_K)}$ with ${\nu_j=n_j/N}$.
Here, the square root of the determinant of the expected value of the dyadic
square of the $p$-gradient of $G$ is a factor in the prior
density~\cite{note:independence}, 
\begin{equation}\label{eq:4.7}
    (\D\rho)=(\D p)\,\Bigglb|\det{\left\{\overline{{\left(
            \frac{\partial G}{\partial p_j}\frac{\partial G}{\partial p_k}
            \right)}_{jk}}\right\}}\Biggrb|^{1/2}
            w_{\mathrm{cstr}}(p)\,,
\end{equation}
where $\overline{\,f(\nu)\,}$ denotes the expected value of $f(\nu)$,
\begin{equation}\label{eq:4.8}
  \overline{\,f(\nu)\,}=\sum_D L(D|\rho)f(\nu)\,.
\end{equation}
We have, in particular, the generating function
\begin{equation}\label{eq:4.9}
  \overline{\,\exp{\left(\sum_{k=1}^Ka_k\nu_k\right)}\,}
=\left(\sum_{k=1}^K\Exp{a_k/N}p_k\right)^N
\end{equation}
for the expected values of products of the $\nu_k$s.
The prior defined by Eq.~(\ref{eq:4.7}) is form-invariant in the same sense,
and for the same reason, as the prior of Eq.~(\ref{eq:4.5}).

\begin{table}
\caption{\label{tbl:priors}%
Form-invariant priors constructed by one of the two methods described in the
text.
The ``$\,\sqrt{\det}\,$'' column gives the $p$-dependent factors only 
and omits all $p$-independent constants.
The first method [Eq.~(\ref{eq:4.5})] proceeds from functions of the
probabilities that have extremal values when all probabilities are equal or
all vanish save one. 
The second method [Eq.~(\ref{eq:4.7})] uses functions that quantify how
similar are the probabilities and the frequencies.
The ``hedged prior'' is named in analogy to the 
``hedged likelihood''~\cite{Blume-Kohout:10}. }
\centering
\begin{ruledtabular}
\begin{tabular}{@{}ccc@{}}
method & primary function & $\sqrt{\det}$ \\ 
\hline
\rule{0pt}{14pt}1st 
& $\ds-\sum_kp_k\log p_k$ & $\ds\frac{1}{\sqrt{p_1p_2\cdots p_K}}$ \\
    & (Shannon entropy) & (Jeffreys prior)\\[2ex]
1st & $\ds\sum_kp_k^2$ & 1 \\ & (purity) & (primitive prior)\\[2ex]
2nd & $\ds\sum_k\nu_kp_k$ &  $\sqrt{p_1p_2\cdots p_K}$\\
    & (inner product)  & (hedged prior)\\[2ex]
2nd & $\ds\sum_k\nu_k\log(\nu_k/p_k)$ & $\ds\frac{1}{\sqrt{p_1p_2\cdots p_K}}$
\\  & (relative entropy)  & (Jeffreys prior)
\end{tabular}
\end{ruledtabular}
\end{table}

Table~\ref{tbl:priors} reports a few examples of ``$\,\sqrt{\det}\,$'' factors
constructed by one of these two methods.  
It is worth noting that the Jeffreys prior can be obtained from the entropy of
the probabilities by the first method as well as from the relative entropy
between the probabilities and the frequencies by the second method. 
The latter is a variant of Jeffreys's original derivation~\cite{Jeffreys:46}
in terms of the Fisher information.

\subsection{Conjugation}\label{sec:conjugate}
Sometimes there are reasons to expect that the actual state is close to a
certain target state with probabilities ${t=(t_1,t_2,\ldots,t_K)}$.
This is the situation, for example, when a source is designed to emit the
quantum-information carriers in a particular state.
A \emph{conjugate prior}
\begin{equation}\label{eq:4.10}
  (\D\rho)=(\D p)\left(p_1^{t_1}p_2^{t_2}\cdots p_K^{t_K}\right)^{\alpha}
           w_{\mathrm{cstr}}(p)
\quad\mbox{with $\alpha>0$}
\end{equation}
could then be a natural choice~\cite{note:conjugate}.
The $(\cdots)^{\alpha}$ factor is maximal for ${p=t}$, and the peak is
narrower when $\alpha$ is larger.

The conjugate prior can be understood as the ``mock posterior'' for the
primitive prior that results from pretending that $\alpha$ copies have been
measured in the past and data obtained that are most typical for the target
state. 
Therefore, a conjugate prior is quite a natural way of expressing the
expectation that the apparatus is functioning well.
The posterior content of a region will be data-dominated only if $N$ is much
larger than $\alpha$.

In this context, it may be worth noting that the Bayesian mean state,
\begin{equation}\label{eq:4.11}
  \rhoBM=\int\limits_{\cR_0}(\D\rho)\,\rho\,,
\end{equation}
computed with the conjugate prior above, is usually not the target state
unless $\alpha$ is large. 
One could construct priors for which $\rhoBM$ is the target state,
but the presence of the $w_{\mathrm{cstr}}(p)$ factor requires a case-by-case
construction.

\subsection{Marginalization}\label{sec:margin}
All priors used as examples---the ones in Table~\ref{tbl:priors} and
Eqs.~(\ref{eq:4.2}), (\ref{eq:4.4}), (\ref{eq:4.10})---have in common that
they are defined in terms of the probabilities and, therefore, they refer to
the particular POM with which the data are collected.
While this pays due account to the significance of the data, it does not seem
to square with the point of view that prior probabilities are solely a property
of the physical processes that put the quantum-information carriers into the
state that is then diagnosed by the POM.

When adopting this viewpoint, one begins with a prior density defined on the
entire state space.
In addition to the parameters that specify the reconstruction space (essentially
the probabilities $p$), this full-space prior will depend on parameters whose
values are not determined by the data.
There could be very many nuisance parameters of this kind, as
illustrated by the somewhat extreme harmonic-oscillator example of
Sec.~\ref{sec:ReconSp}. 
Upon integrating the full-space prior over the nuisance parameters, one
obtains a \emph{marginal prior} on the reconstruction space.
As a function on the reconstruction space, 
the marginal prior is naturally parameterized in terms of the probabilities
and so fits into the formalism we are using throughout.

Harking back to the last paragraph in Sec.~\ref{sec:ReconSp}, we note that the
invoking of ``additional criteria or principles'' is exactly what would be
required if one wishes to report estimated values of the nuisance parameters.
That, however, goes beyond making statements that are solidly supported by
the data and is, therefore, outside the scope of this article.   

\begin{figure}
\centering
\includegraphics{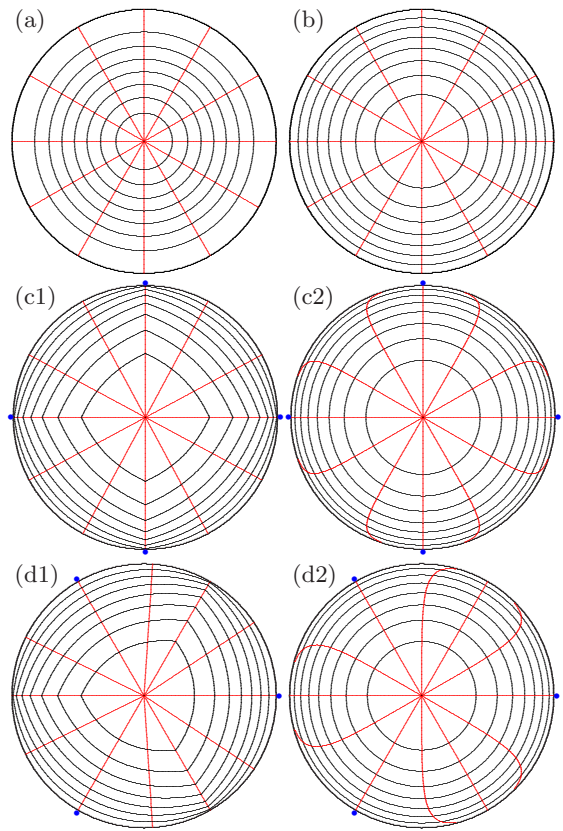}
\caption{\label{fig:tilings}%
Uniform tilings of the unit disk for four different priors.
The disk is in the $xy$ plane, with the $x$ axis horizontal, the $y$ axis
vertical, and the disk center at ${x=y=0}$. 
Tiling (a) is for the marginal prior of Eq.~(\ref{eq:4.12});
tiling (b) depicts the primitive prior of Eq.~(\ref{eq:5.6}); 
tilings (c1) and (c2) illustrate the Jeffreys prior of Eq.~(\ref{eq:5.7})
with the blue dots ({\color{blue}\scriptsize$\bullet$}) just outside the unit
circle indicating the four directions onto which the POM outcomes project;
and tilings (d1)and (d2) are for the Jeffreys prior of Eq.~(\ref{eq:5.8}), the
blue dots marking the three directions of the trine projectors.
In each tiling, we identify $96$ regions of equal size by dividing the disk
into eight ``tree rings'' of equal size and twelve ``pie slices'' of equal size.
In the tilings (a), (b), (c1), and (d1), the boundaries of the pie slices are
(red) rays and an arc of the unit circle; in the tilings (a), (b), (c2), and
(d2), the tree rings have concentric circles as their boundaries.
}
\end{figure}

The symmetric uniform-in-purity prior of Secs.~\ref{sec:utility} and
\ref{sec:symmetric} provides an example for marginalization if the POM only
gives information about ${x=\langle\sigma_x\rangle}$ and
${y=\langle\sigma_y\rangle}$ but not about ${z=\langle\sigma_z\rangle}$.
We express the full-space prior in cartesian coordinates, integrate over
$z$, and arrive at
\begin{eqnarray}\label{eq:4.12}
  (\D\rho)&=&\D x\,\D y\,\frac{1}{2\pi}\int\limits_{-\infty}^{\infty}\D z\,
\frac{\eta(1-x^2-y^2-z^2)}{\sqrt{x^2+y^2+z^2}}
\nonumber\\ &=&
\D x\,\D y\,\frac{1}{\pi}\eta(1-x^2-y^2)
\cosh^{-1}\frac{1}{\sqrt{x^2+y^2}}\,.\qquad
\end{eqnarray}
This marginal prior is a function on the unit disk in the $xy$ plane, which is
the natural choice of reconstruction space here.
When one expresses $(\D\rho)$ in polar coordinates, 
${x+\I y=s\Exp{\I\varphi}}$ with ${s\geq0}$, one sees that $(\D\rho)$ is
uniform in $\varphi$ and 
in ${s^2\cosh^{-1}(1/s)-\sqrt{1-s^2}}$, which increases monotonically from $-1$
to $0$ on the way from the center of the disk at ${s=0}$ to the unit circle
where ${s=1}$.
Plot (a) in Fig.~\ref{fig:tilings} illustrates the matter.

\section{Examples}\label{sec:examples}
For illustration, we consider the simplest situation that exhibits the typical
features:
The quantum-information carriers have a qubit degree of freedom, which is
measured by one of two standard POMs that are not informationally complete.

\subsection{POMs and priors}\label{sec:ex-POMpriors}
For both POMs, the unit disk in the $xy$ plane suggests itself for the
reconstruction space $\cR_0$.  
The first POM combines projective measurements of $\sigma_x$ and $\sigma_y$
into a four-outcome POM (${K=4}$) with probabilities
\begin{equation}\label{eq:5.1}
  \begin{array}{l}p_1\\p_2\end{array}\biggr\}
   =\frac{1}{4}(1\pm x)\,,\quad
  \begin{array}{l}p_3\\p_4\end{array}\biggr\}
   =\frac{1}{4}(1\pm y)\,.
\end{equation}
The permissible probabilities are identified by
\begin{equation}\label{eq:5.2}
 w_{\mathrm{cstr}}(p)\dot{=}\eta(p)\,\delta(p_1+p_2-\tfrac{1}{2})\,
                   \delta(p_3+p_4-\tfrac{1}{2})\,\eta(3-8p^2) \,,
\end{equation}
where
\begin{equation}\label{eq:5.3}
  \eta(p)=\prod_{k=1}^K\eta(p_k)\quad\mbox{and}\quad
  p^2=\sum_{k=1}^Kp_k^2\,.
\end{equation}
The dotted equal sign in Eq.~(\ref{eq:5.2}) stands for ``equal up to a
multiplicative constant,'' namely the factor that ensures the unit size of the
reconstruction space.

The second POM is the three-outcome trine measurement (${K=3}$), whose
outcomes are  subnormalized projectors on the eigenstates of $\sigma_x$ and
${(-\sigma_x\pm\sqrt{3}\,\sigma_y)/2}$ with eigenvalue~$+1$. 
It has the
probabilities
\begin{equation}\label{eq:5.4}
  p_1=\frac{1}{3}(1+x)\,,\quad
  \begin{array}{l}p_2\\p_3\end{array}\biggr\}
   =\frac{1}{6}(2-x\pm\sqrt{3}\,y)\,,
\end{equation}
for which
\begin{eqnarray}\label{eq:5.5}
 w_{\mathrm{cstr}}(p)\dot{=}\eta(p)\,\delta(p_1+p_2+p_3-1)\,
                   \eta(1-2p^2) 
\end{eqnarray}
summarizes the constraints that the permissible values of $p_1$, $p_2$, $p_3$
obey. 

Both POMs have the same primitive prior,
\begin{equation}\label{eq:5.6}
  (\D\rho)=\D x\,\D y\,\frac{1}{\pi}\eta(1-x^2-y^2)
          =\D s^2\,\frac{\D\varphi}{2\pi}\,,
\end{equation}
where ${0\leq s\leq1}$ and $\varphi$ covers any convenient range of $2\pi$.
This prior is uniform in $x$ and $y$, and in $s^2$ and $\varphi$.
The polar-coordinate version is the more natural parameterization of the unit
disk; it is used for plot (b) in Fig.~\ref{fig:tilings}.

The Jeffreys prior for the four-outcome POM is~\cite{note:normalization}
\begin{equation}\label{eq:5.7}
  (\D\rho)\dot{=}\frac{\D s\,s\,\D\varphi}
                      {\sqrt{1-s^2+\tfrac{1}{4}s^4\sin(2\varphi)^2}}\,.
\end{equation}
Plots (c1) and (c2) in Fig.~\ref{fig:tilings} show uniform tilings
of the unit disk for this prior.
For the three-outcome POM, we have the Jeffreys prior~\cite{note:normalization}
\begin{equation}\label{eq:5.8}
  (\D\rho)\dot{=}\frac{\D s\,s\,\D\varphi}
                      {\sqrt{1-\tfrac{3}{4}s^2+\tfrac{1}{4}s^3\cos(3\varphi)}} 
\end{equation}
and the tilings of plots (d1) and (d2) in Fig.~\ref{fig:tilings}.
The cross-hairs symmetry of the four-outcome POM and the trine symmetry of the
three-outcome POM are manifest in their respective uniform tilings.

\begin{figure}
\centering%
\includegraphics{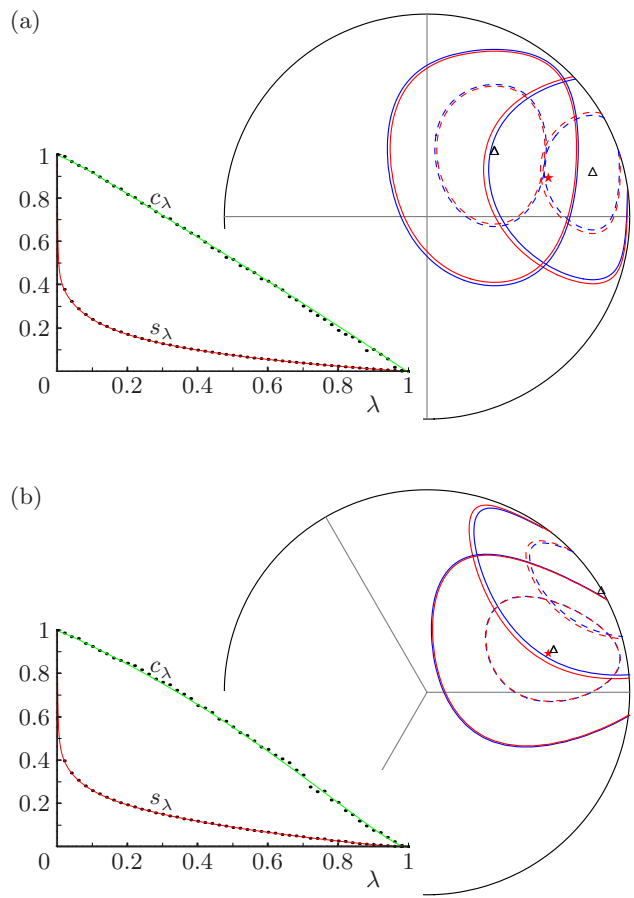}
\caption{\label{fig:regions}
Smallest credible regions for simulated experiments. 
Twenty-four copies are measured by the POMs of Sec.~\ref{sec:ex-POMpriors}, 
which have the unit disk 
of Fig.~\ref{fig:tilings} as the reconstruction space.
Plot~(a) is for the four-outcome POM with the cross hairs indicating the
orientations of the two projective measurements.
Plot~(b) is for the three-outcome measurement with the orientation of the
trine indicated.
The red star ({\color{red}$\star$}) at ${(x,y)=(0.6,0.2)}$ marks the actual
state that was used for the simulation. 
For each POM, there are SCRs for the data of two simulated experiments,
with black triangles ({\scriptsize$\triangle$}) indicating the respective MLEs.
The boundaries of the SCRs with credibility ${c=0.9}$ are traced by the
continuous lines; all of these SCRs contain the actual state. 
The dashed lines are the boundaries of the SCRs with
credibility ${c=0.5}$; the actual state is inside half of these SCRs.
Red lines are for the primitive prior of Eq.~(\ref{eq:5.6}), the blue lines
are for the Jeffreys priors of Eqs.~(\ref{eq:5.7}) and (\ref{eq:5.8}),
respectively. ---
The insets in the lower left corners show the size $s_\lambda$ and the
credibility $c_\lambda$ for the BLRs of two simulated experiments.
Inset (a) is for $(6,3,10,5)$ counts for the four-outcome POM and the Jeffreys
prior; inset (b) is for $(13,7,4)$ counts for the three-outcome POM and the
primitive prior. 
The dots show the values computed with a Monte Carlo algorithm. 
There is much more scatter in the $c_\lambda$ values than the $s_\lambda$
values.
The red lines are fits to the $s_\lambda$ values, with the fits using twice as
many values than there are dots in the insets. 
The green lines that approximate the $c_\lambda$ values
are obtained from the red lines with the aid of Eq.~(\ref{eq:3.12a}).
}
\end{figure}

\subsection{Simulated measurements}\label{sec:ex-simulations}
Figures~\ref{fig:regions}(a) and \ref{fig:regions}(b) show SCRs obtained for
simulated experiments in which ${N=24}$ copies of a qubit state are measured.
The actual state used for the simulation has ${x=0.6}$ and ${y=0.2}$.
Its position in the reconstruction space is indicated by the red star
({\color{red}$\star$}). 

In Fig.~\ref{fig:regions}(a), we see the SCRs for the four-out\-come POM.
Two measurements were simulated, with ${(n_1,n_2,n_3,n_4)=(8,5,10,1)}$ and
${(6,3,10,5)}$ clicks of the detectors, respectively, and the triangles
({\scriptsize$\triangle$}) show the positions of the corresponding MLEs.
For each data, the plot reports the SCRs with credibility ${c=0.5}$ and
${c=0.9}$, both for the primitive prior of Eq.~(\ref{eq:5.6}) and for the
Jeffreys prior of Eq.~(\ref{eq:5.7}).
The actual state is inside two of the four SCRs with
credibility ${c=0.5}$ and is contained in all four SCRs with credibility
${c=0.9}$. 

Not unexpectedly, we get quite different regions for the two rather different
sets of detector click counts.
Yet, we observe that the choice of prior has little effect on the SCRs,
although the total number of measured copies is too small for relying on the
consistency of the priors.
The same remarks apply to the SCRs for the three-outcome POM in
Fig.~\ref{fig:regions}(b); here we counted ${(n_1,n_2,n_3)=(15,8,1)}$ and
$(13,7,4)$ detector clicks in the simulated experiments.

In Sec.~\ref{sec:scBLR} we remarked that the estimator regions are properly
communicated by reporting $s_\lambda$ and $c_\lambda$ as functions of
$\lambda$.
This is accomplished by the insets in Fig.~\ref{fig:regions} for two of the
four simulated experiments. 
The dots give the values obtained by numerical integration that uses a Monte
Carlo algorithm. 
The scatter of these numerical values confirms the expected:
The computation of $s_\lambda$ only requires sampling the probability space
in accordance with the prior and determining the fraction of the sample that
is in $\cR_\lambda$; for the computation of $c_\lambda$ we need to add the
values of $L(D|\rho)$ for the sample points inside $\cR_\lambda$; 
and since $L(D|\rho)$ is a sharply peaked function of the probabilities, 
the $s_\lambda$ values are more trustworthy than the $c_\lambda$ values 
for the same computational effort.  
The line fitted to the $s_\lambda$ values is a Pad\'e approximant (see, for
example, section 5.12 in Ref.~\cite{NR}) 
that takes the analytic forms near ${\lambda=\lambda_0=0}$ and ${\lambda=1}$
into account.  
The line approximating the $c_\lambda$ values is then computed in accordance
with Eq.~(\ref{eq:3.12a}).

\section{Outlook}\label{sec:outlook}
For the given data and chosen credibility, the SCR is a neighborhood of the
MLE. In this sense, then, one can regard the SCR as identifying error bars on
the parameter values of the MLE in a systematic way.
Thereby, the MLE is often a state whose probabilities equal the
observed frequencies, and if there is no such state in the reconstruction space,
efficient methods are at hand for computing the MLE.
We are, however, currently lacking equally efficient algorithms for finding
the SCR. 

Progress on this front is needed before one can apply the concepts of MLRs and
SCRs to situations in which the reconstruction space is of high dimension.
Upon recalling that informationally complete POMs for two-qubit systems
already have a 15-dimensional reconstruction space, the need for powerful
numerical schemes is utterly plain.

In many applications, one is interested in a few parameters only, perhaps a
single one, such as the concurrence of a two-qubit state or its fidelity with
a target state.
It may then be possible to reduce the dimensionality of the problem by
marginalizing the nuisance parameters, preferably proceeding from a
utility-based prior. 

Even after such a reduction, there remains the challenge of evaluating the
multi-dimensional integrals that tell us the size of the BLRs, and then their
credibility, so that we can identify the looked-for MLR and SCR.
For this purpose one needs good sampling strategies~\cite{note:sampling}.
It is suggestive to rely on the data themselves for guidance.
The full sequence of detector clicks identifies the MLE of the data, and
subsequences---chosen randomly or systematically---have their own MLEs.
These boot-strapped MLEs are expected to accumulate in the vicinity of the
full-data MLE and may so provide a useful sampling method.
We have just begun to enter this unexplored territory and will report progress
in due course.

We close with a general observation.
MLEs, MLRs, SCRs, and confidence regions are concepts of statistics,
even if the terminology is not universal.
As we have seen, the quantum aspect of the state estimation problem enters
only through the Born
rule which restricts the probabilities to those obtainable from a POM and a
bona fide statistical operator.
Except for these restrictions, there is no difference between state estimation
in quantum mechanics and standard statistics.
Accordingly, quantum mechanicians can benefit much from the methods
developed by statisticians.

\acknowledgments
We benefitted greatly from discussions with David Nott and thank him in
particular for bringing Refs.~\cite{Evans+2:06} and \cite{Kass+1:96} to our
attention. 
The Centre for Quantum Technologies is a Research Centre
of Excellence funded by the Ministry of Education and the National
Research Foundation of Singapore.


\begin{thebibliography}{00}
\bibitem{note:typical}
It is advisable to verify that the observed sequence does not have systematic
correlations that speak against the assumption of independently and
identically prepared quantum-information carriers.
\bibitem{note:eta-delta}
The symbol $\eta(\ )$ denotes Heaviside's
unit step function, and $\delta(\ )$ is Dirac's delta function.
\bibitem{note:imperfections}
It is possible to account for detector inefficiencies (some carriers escape
detection) and dark counts (spontaneous detector clicks), but such technical
details, as important they may be in practical applications, are not material
to the current discussion.
\bibitem{note:concave}
The negative logarithm of the point likelihood is $N$ times the sum of the
relative entropy between the probabilities $p$ and the frequencies $\nu$, and
the Shannon entropy of the frequencies (see Table~\ref{tbl:priors}),
\begin{displaymath}
  -\frac{1}{N}\log L(D|\rho)=\sum_k\nu_k\log\frac{\nu_k}{p_k}
                            -\sum_k\nu_k\log\nu_k\,.
\end{displaymath}
Since the relative entropy is a convex function of the probabilities, the
logarithm of the point likelihood is a concave function of $p$.
\bibitem{note:evidence}
If we wish to be quantitative about these beliefs,
we can use the number $10\log_{10}(C_{\cR}/S_{\cR})$ to
measure the evidence for the hypothesis that the actual state
is in region $\cR$ (in units of dB). 
Then there is more evidence in favor of the BLR $\cR_{\lambda}$ than for
any other region of the same credibility.
\bibitem{note:independence}
Since $w_{\mathrm{cstr}}(p)$ enforces all constraints, the $p_k$s are
independent variables when $F(p)$ and $G(p,\nu)$ are differentiated in
Eqs.~(\ref{eq:4.5}) and (\ref{eq:4.7}), respectively.
\bibitem{note:conjugate}
Such priors are called ``conjugate'' in standard statistics literature
because the $(\cdots)^{\alpha}$ factor has the same structure as the point
likelihood: a product of powers of the detection probabilities.
\end{thebibliography}
\end{document}